\documentclass[10pt,conference]{IEEEtran}
\IEEEoverridecommandlockouts
\usepackage{cite}
 \usepackage{color}

\usepackage{float}
\usepackage{amsmath,amssymb,amsfonts}
\usepackage{algorithmic}
\usepackage{graphicx}
\usepackage[left=1.35cm,right=1.35cm,top=1.9cm,bottom=4.23cm]{geometry}
\usepackage{textcomp}
\usepackage{xcolor}
\usepackage{bbm}
\def\BibTeX{{\rm B\kern-.05em{\sc i\kern-.025em b}\kern-.08em
    T\kern-.1667em\lower.7ex\hbox{E}\kern-.125emX}}
\usepackage[subtle,tracking=normal]{savetrees}
\begin{document}
\def\blue{\textcolor{blue}}
\title{Accelerating Graph Neural Networks via Edge Pruning for Power Allocation in Wireless Networks\\

}
\newcommand\blfootnote[1]{%
  \begingroup
  \renewcommand\thefootnote{}\footnote{#1}%
  \addtocounter{footnote}{-1}%
  \endgroup
}
\author{\IEEEauthorblockN{Lili Chen, Jingge Zhu and Jamie Evans}
\IEEEauthorblockA{Department of Electrical and Electronic Engineering, University of Melbourne, Australia \\
Email: lilic@student.unimelb.edu.au, jingge.zhu@unimelb.edu.au, jse@unimelb.edu.au\\
}

}

\maketitle

\begin{abstract}
Graph Neural Networks (GNNs) have recently emerged as a promising approach to tackling power allocation problems in wireless networks. Since unpaired transmitters and receivers are often spatially distant, the distance-based threshold is proposed to reduce the computation time by excluding or including the channel state information in GNNs. In this paper, we are the first to introduce a neighbour-based threshold approach to GNNs to reduce the time complexity. Furthermore, we conduct a comprehensive analysis of both distance-based and neighbour-based thresholds and provide recommendations for selecting the appropriate value in different communication channel scenarios. We design the corresponding neighbour-based Graph Neural Networks (N-GNN) with the aim of allocating transmit powers to maximise the network throughput. Our results show that our proposed N-GNN offer significant advantages in terms of reducing time complexity while preserving strong performance and generalisation capacity. Besides, we show that by choosing a suitable threshold, the time complexity is reduced from $\mathcal{O}(|\mathcal{V}|^2)$ to $\mathcal{O}(|\mathcal{V}|)$, where $|\mathcal{V}|$ is the total number of transceiver pairs.

\end{abstract}


\section{Introduction}\label{sec:introduction}
The proliferation of fifth-generation (5G) communication technology has resulted in an escalating need for high-rate wireless access services. This trend has engendered formidable challenges to the availability of spectrum resources due to their finite nature. As an auspicious technique to alleviate the predicament of spectrum scarcity, device-to-device (D2D) communications have recently garnered considerable attention\cite{agiwal2016next}.

In D2D communication, two devices can communicate directly without relying on the involvement of the base station (BS) or access point. The short-range D2D links facilitate high data rates for local devices, reduce power consumption in mobile devices, and relieve the traffic of BSs. However, power allocation in D2D networks is often a non-convex problem and computationally hard. Inspired by recent success in computer science, graph neural networks (GNNs) have been applied to power allocation problems in wireless networks. 
\blfootnote{The work was supported by the Melbourne Research Scholarship of the University of Melbourne and in part by the Australian Research Council under projects DE210101497 and DP220103281.}
In D2D networks, unpaired transmitters and receivers are often spatially distant. In light of the distance-dependent nature of interference decay, eliminating connections between distant nodes appears to be a logical approach towards reducing computational complexity, while not significantly compromising performance.
 
In \cite{shen2020graph}, the authors proposed the distance-based threshold to reduce computation time by excluding the channel state information (CSI) in GNN when the distance is over a specific threshold. Similarly, the authors in \cite{wang2022learning} used the channel-based threshold to reduce the complexity by removing the edges between the transmitter to the non-paired receivers if the CSI is smaller than the threshold.

However, the threshold was chosen arbitrarily, lacking further elaboration. In response to this issue, a recent study explored the delicate balance between performance and time complexity by implementing the distance-based threshold \cite{chen2023graph}. The authors demonstrated that selecting an appropriate threshold can significantly reduce the anticipated time complexity, while concurrently preserving favorable performance outcomes.

However, the results presented in \cite{chen2023graph} are restricted to a single path loss exponent (PLE), therefore, their generalisability to alternative path loss exponents is not assured. To address this limitation, the present investigation undertakes a comprehensive analysis of thresholds and provides recommendations for selecting the appropriate threshold value that corresponds to different PLEs. To accomplish this objective, we evaluate various thresholds in terms of their potential to catch the expected interference. Furthermore, we introduce a neighbour-based threshold that enables the reduction of time complexity.

The contributions of this paper are summarised as follows:
\begin{itemize}
    \item This research is the first to introduce a neighbour-based threshold approach to GNNs that offers significant advantages in terms of reducing time complexity while preserving strong performance. 
    \item This study is the first to systematically investigate appropriate threshold selection from a stochastic geometry perspective. We provide recommendations for selecting the appropriate threshold value in terms of their potential to catch the expected interference. Our findings highlight the importance of carefully considering these thresholds and the potential implications of their selection, which can vary based on the specific wireless networks in which they are applied.

    \item We conduct extensive experiments to verify the effectiveness of our proposed guideline for selecting the appropriate threshold. We demonstrate the neighbour-based threshold is preferable under the network density of interest. We show that by choosing a suitable neighbour-based threshold, the strong performance and generalisation capacity are preserved, while the time complexity is reduced from $\mathcal{O}(|\mathcal{V}|^2)$ to $\mathcal{O}(|\mathcal{V}|)$, where $|\mathcal{V}|$ is the total number of transceiver pairs.

\end{itemize}

\section{Preliminaries}

\subsection{System Model}\label{sub:systemmodel}
We consider a wireless communication network containing $T$ transmitters, where they all share the same channel spectrum. We denote the index set for transmitters by $\mathcal{T} = \{1,2,...,T\}$. For each $t$ in $\mathcal{T}$, we define $\mathrm{D}(t)$ to be the index of the paired receiver. The received signal at the $\mathrm{D}(t)$-th receiver for any $t \in \mathcal{T}$ is given by
\begin{equation}
y_{\mathrm{D}(t)}=h_{t, \mathrm{D}(t)} s_{t}+\sum_{j\in  \mathcal{T} \backslash\{t\}} h_{ j,\mathrm{D}(t)} s_{j}+ n_{\mathrm{D}(t)},\quad t \in \mathcal{T},
\end{equation}
where $h_{t,\mathrm{D}(t)} \in \mathbb{C}$ represents the communication channel between $t$-th transmitter and its intended $\mathrm{D}(t)$-th receiver, $h_{ j,\mathrm{D}(t)} \in \mathbb{C}$ represents the interference channel between $j$-th transmitter and $\mathrm{D}(t)$-th receiver. 

The transmitted data symbol for the $t$-th transmitter is represented by $s_{t} \in \mathbb{C}$, while the additive Gaussian noise at the $\mathrm{D}(t)$-th receiver is modeled as $n_{\mathrm{D}(t)} \sim \mathcal{C} \mathcal{N}\left(0, \sigma_{\mathrm{D}(t)}^{2}\right)$. The signal-to-interference-plus-noise ratio (SINR) of the $\mathrm{D}(t)$-th receiver is expressed as follows:
\begin{equation}
\operatorname{SINR}_{\mathrm{D}(t)}=\frac{\left|h_{t,\mathrm{D}(t)}\right|^{2} p_{t}}{ \sum_{j\in  \mathcal{T} \backslash\{t\}} \left|h_{j, \mathrm{D}(t)}\right|^{2} p_{j}  +\sigma_{\mathrm{D}(t)}^{2}}, \quad  t \in \mathcal{T},
\end{equation}
where $p_{t}=\mathbb{E}\left[|s_{t}|^{2}\right]$ is the power of the $t$-th transmitter. 

We denote $\mathbf{p}=\left[p_{1}, \cdots,    p_{T}\right]$ as the power allocation vector. For a given power allocation vector $\mathbf{p}$ and channel information $\left\{h_{i j}\right\}_{i \in \mathcal{T}, j \in \mathrm{D}(i)}$, the achievable rate $\mathcal{R}_{\mathrm{D}(t)}$ of the $\mathrm{D}(t)$-th receiver is given by
\begin{equation}
\mathcal{R}_{\mathrm{D}(t)}(\mathbf{p})=\log_2 \left(1+\operatorname{SINR}_{\mathrm{D}(t)} \right), \quad t \in \mathcal{T}.
\end{equation}

\subsection{Optimisation Problem}

In this study, we address the problem of maximising the weighted sum-rate, a widely studied optimisation problem in the literature \cite{shen2020graph, chen2023graph, shi2011iteratively, liang2019towards}. The objective is to maximise the performance under maximum power constraints, which is formulated as,
 \begin{equation}
    \begin{array}{cl}
    \underset{\mathbf{p}}{\operatorname{maximise}} & \sum_{t} w_{t}  \mathcal{R}_{\mathrm{D}(t)}(\mathbf{p}), \\
    \text { subject to } & 0 \leq p_{t} \leq P_{\max }, \quad  \forall t \in \mathcal{T},
    \end{array}
    \label{eq:weightsumrate}
\end{equation}   
where $w_{t} \in [0,1]$ is the weight for the ${t}$-th transmitter and $P_{\max}$ is the maximum power constraint for transmitters. 

\section{Threshold-based Graph Neural networks for power allocation}
In this section, we propose a general guideline for selecting appropriate thresholds that achieves reasonably good performance while minimising complexity. Specifically, we provide recommendations in terms of their potential to catch the expected interference. We then formulate the proposed distance-based and neighbour-based graph representations and apply them to the GNNs for power allocation problems.
\subsection{Traditional Graph Representation}

\begin{figure}[htbp]
\centerline{\includegraphics[width=6cm]{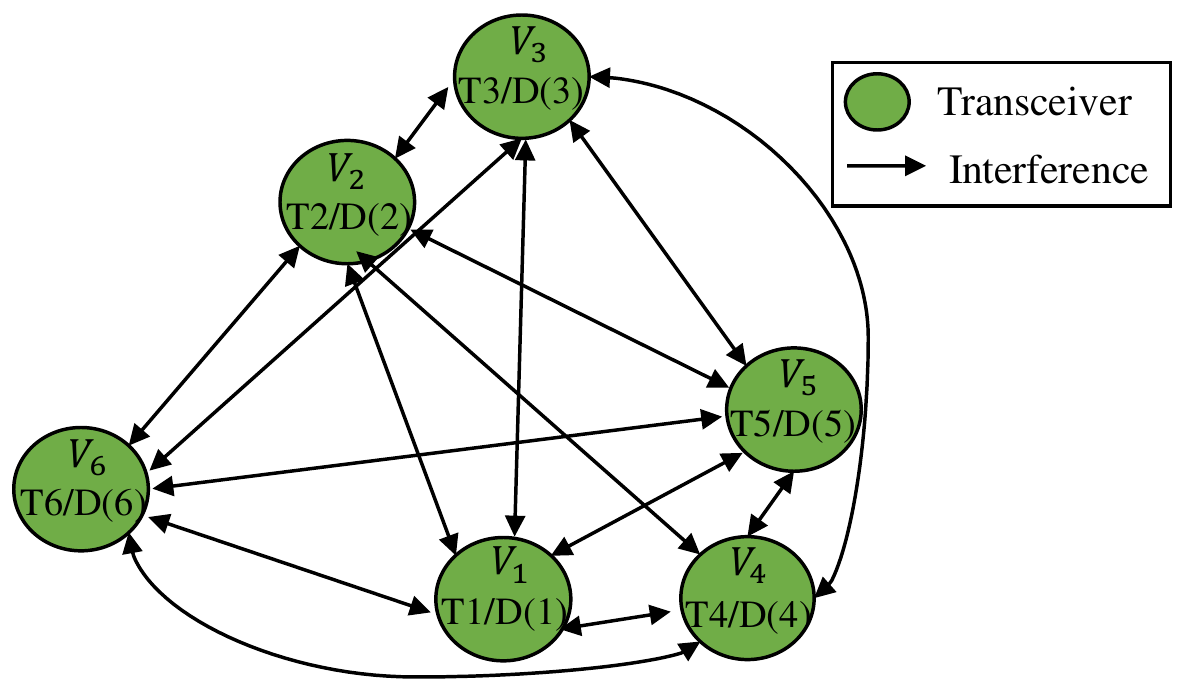}}
\caption{D2D communication network.}
\label{fig:D2Dfullfull}
\end{figure}

The most common graph representation for power allocation in D2D networks is treating the transceiver pairs as a vertex\cite{shen2020graph}. The edge between the transceiver pairs can represent the interference between them (see Figure~\ref{fig:D2Dfullfull} as an example). Normally, this graph representation is a complete graph due to the existing interference between each transceiver pair. However, since unpaired transmitters and receivers are often spatially distant, this pair might have negligible effects on power allocation due to interference decay with distance. Besides, this complete graph would induce relatively high complexity since the complexity of GNNs depends on the total number of edges\cite{shen2020graph}.

\subsection{The Proposed Guideline for Choosing Appropriate Threshold}
To address these issues, the edges can be excluded based on the threshold. Appropriate threshold selection is typically achieved by running algorithms for various densities and PLEs. However, this approach is time-consuming due to the dynamic nature of wireless network environments. To address this challenge, we aim to find a good proxy for the appropriate threshold. We first derive a mathematical expression that models the relationship between the threshold and interference. Generally, we expect that the performance of the GNN will improve if it can capture more information in the network. However, increasing the amount of information captured also results in higher time complexity for the algorithm. To strike a balance between performance and computational efficiency, we hypothesise that if the algorithm can capture a significant proportion of the interference (e.g. $\boldsymbol{95\%}$), then the resulting performance will be satisfactory for most practical applications.

Let us consider an infinite network, we randomly place transceiver pairs $i$ to be the points of a Stationary Poisson Process $\Phi \subset \mathbb{R}^2 $ of intensity $\lambda$\cite{blaszczyszyn2018stochastic}. Denote $r_i$ as the distance between $i$-th transceiver pair and the target pair.  Assuming each transceiver pair is transmitting with unit power and the CSI is $g(r) = \min\{1,(\frac{r}{d_0})^{-\alpha}\}$ with PLE $\alpha$ and reference distance $d_0$. Here, we assume $\alpha >2$ to ensure the expected interference is finite \cite{haenggi2005distances}. According to Campbell's Theorem \cite{blaszczyszyn2018stochastic}, the expected interference for the target pair is
\begin{equation}
\begin{aligned}
E[I]= & E[\sum_{i \in \Phi} g(r_i)] =\lambda \int_0^{2 \pi} \int_0^{\infty} g(r) r d r d \beta \\
 = &\pi \lambda d_0^2+2 \pi \lambda \cdot \frac{d_0^2}{\alpha-2} = \pi \lambda d_0^{2}(1+\frac{2}{\alpha-2})
\end{aligned}
\end{equation}
where $dr$ and $d\beta$ represent differential elements pertaining to the radial and angular directions, respectively.

\subsubsection{Distance-based Threshold}

Consider removing all the transceiver pairs $i$ when the distance $d_i \geq t $, where $t$ is the distance-based threshold. Assuming the threshold $t \geq d_0$, the expected interference $E[I_{d}(t)]$ within this area with threshold $t$ is

\begin{equation}
\begin{aligned}
E\left[I_{d}(t)\right] = \pi \lambda d_0^{2}(1+2\frac{1-(\frac{t}{d_0})^{2-\alpha}}{\alpha-2})
\end{aligned}
\end{equation}

The interference ratio $A_t$ is defined as the expected interference resulting from applying a distance-based threshold $t$ to the expected total interference. The expression is given by,
\begin{equation}
A_{t} = \frac{E[I_{d}(t)]}{E[I]}= \frac{\alpha-2(\frac{t}{d_0})^{2-\alpha}}{\alpha}
\end{equation}

\subsubsection{Neighbour-based Threshold}

For a specific transceiver pair, the probability density function (pdf) of the nearest neighbour within distance $r$ is given by \cite{haenggi2005distances}
\begin{equation}f_{R_1}(r)=2 \lambda \pi r e ^{-\lambda \pi r^{2}}.\end{equation}
Therefore, the expected interference with only the closest neighbour is

\begin{equation}
\begin{aligned}
E\left[I_n(1)\right] & =\int_0^{\infty}  g(r)  f_{R_1}(r) d r  =\int_0^{\infty}  g(r)  2 \lambda \pi r e^{-\lambda \pi r^2} d r \\
& =1-e^{-\lambda \pi d_0^2}+d_0^\alpha \int_{\lambda \pi d_0^2}^{\infty} r^{-\alpha} e^{-t} d t \\
& =1-e^{-\lambda \pi d_0^2}+d_0^\alpha(\lambda \pi)^{\frac{\alpha}{2}}  \Gamma \left(1-\frac{\alpha}{2}, \lambda \pi d_0^2\right)
\end{aligned}
\end{equation}
where $\Gamma(s,x) = \int_x^{\infty} t^{s-1}e^{-t}dt$ is a incomplete gamma function. Similarly, the pdf for the distance to the $n$-th  nearest neighbour ($n \geq 1$) is given by\cite{haenggi2005distances}
\begin{equation}
f_{R_{n}}(r)=e^{-\lambda \pi r^{2}} \cdot \frac{2\left(\lambda \pi r^{2}\right)^n}{r (n-1) !} 
\end{equation}
Therefore, the expected interference with only $n$-th  closest neighbour is given by

\begin{equation}
\begin{aligned}
& E\left[I_n(n)\right]  =\int_0^{\infty} g(r) \cdot f_{R_n}(r) d r \\
& =\int_0^{d_0}  e^{-\lambda \pi r^2} \frac{2\left(\lambda \pi r^2\right)^n}{r \cdot \Gamma(n)} d r+\int_{d_0}^{\infty}(\frac{r}{d_0})^{-\alpha} e^{-\lambda \pi r^2} \cdot \frac{2\left(\lambda \pi r^2\right)^n}{r \cdot \Gamma(n)} d r \\
&=1-\left(\sum_{i=0}^{n-1}\frac{\left.(\lambda \pi r^2\right)^i}{i !}\right) e^{-\lambda \pi d_0^2}+\frac{d_0^\alpha(\lambda \pi)^{\frac{\alpha}{2}}}{\Gamma(n)}  \Gamma \left(n-\frac{\alpha}{2}, \lambda \pi d_0^2\right)
\end{aligned}
\end{equation}
We define the interference ratio $O_n$ as the expected interference with $n$  closest neighbours to the expected total interference. The expression is given by, 
\begin{equation}
\centering
O_n = \frac{\sum_{i=1}^{n} E\left[I_n(i)\right] }{\pi \lambda d_0^{2}(1+\frac{2}{\alpha-2})}
\end{equation}

By systematically varying the PLEs and intensities, we are able to determine the corresponding distances and neighbours required to achieve $95\%$ interference, as detailed in  Table~\ref{Tab:DistanceThreshold} and Table~\ref{Tab:expected95}. Here, we consider the practical intensities up to 0.03 \cite{AUSTRALIA} and set 1 meter as the reference distance\cite{maccartney2014omnidirectional}.

\subsubsection{The Proposed Guideline} \label{subsub:proposedguideline}

To evaluate the relative strengths and weaknesses of distance-based and neighbour-based thresholds, we conducted a simulation using a Poisson Point Process with varying intensities. The simulation was performed on a square area with dimensions $B = 10000 m^2$, where the number of transceiver pairs was modelled as a Poisson random variable with mean $\lambda B$ \cite{haenggi2012stochastic}. Once the number of transceiver pairs was determined, they were randomly distributed within the area. 
\begin{table}
\centering
\caption{The expected unit distance required to achieve $90\%$, $95\%$ and $98\%$ of the total interference ($d_0 =1$).}
\begin{tabular}{|l|c|c|c|c|c|c|}
\hline  & $\alpha = 3$ & $\alpha = 3.5 $  & $\alpha = 4 $& $\alpha = 4.5 $  & $\alpha = 5 $ & $\alpha = 5.5 $   \\
\hline$90\% $  & 7 & 4  &  3& 2 &  2& 2 \\
\hline$95\% $  & 12 & 6   & 4 & 3 & 2& 2\\
\hline$98\%  $ & 26  & 10  & 5 & 4 & 3 & 3 \\
\hline
\end{tabular}
\label{Tab:DistanceThreshold}
\end{table}

\begin{table}
    \centering
    \caption{The number of nearest neighbours required to capture $95\%$ of the total interference on average.}
    \begin{tabular}{|l|c|c|c|c|c|c|}
    \hline  & $\alpha = 3$ & $\alpha = 3.5 $  & $\alpha = 4 $& $\alpha = 4.5 $  & $\alpha = 5 $ & $\alpha = 5.5 $  \\
    \hline$\lambda = 0.002 $ & 2 & 1  & 1 & 1 & 1 & 1 \\
    \hline$\lambda = 0.004 $  &  3 & 2  & 1 & 1 & 1 & 1 \\
    \hline$\lambda = 0.01 $   & 5 & 2  & 2 & 1 & 1 & 1 \\
    \hline$\lambda = 0.02 $   & 9 & 3  & 2 & 2 & 2 & 2 \\
    \hline$\lambda = 0.03 $  & 13 & 4  & 2 & 2 & 2 & 2 \\
    \hline
    \end{tabular}
    \label{Tab:expected95}
\end{table}
\begin{table}
\centering
\caption{The variance of the interference for distance-based threshold.}
\begin{tabular}{|l|c|c|c|c|c|c|}
\hline &$\alpha = 3$ & $\alpha = 3.5 $  & $\alpha = 4$ & $\alpha = 4.5 $  & $\alpha = 5 $ & $\alpha = 5.5 $   \\
\hline$\lambda =0.002 $  & 3.07  & 10.01 &13.93 & 25.33 & 150.20& 163.76     \\
\hline$\lambda =0.004 $  & 0.97 &  1.99& 2.83& 5.52 & 21.05&23.12      \\
\hline$\lambda =0.01  $  & 0.25  & 0.39& 0.50 & 0.34 & 1.48& 2.47    \\
\hline$\lambda =0.02  $  & 0.10  & 0.19 &0.19& 0.22 &0.46&0.98   \\
\hline$\lambda =0.03  $  & 0.06 & 0.12 & 0.12& 0.14  &0.28& 0.58     \\
\hline
\end{tabular}
\label{Tab:si_dis_var}
\end{table}
\begin{table}
\centering
\caption{The variance of the interference for neighbour-based threshold.}
\begin{tabular}{|l|c|c|c|c|c|c|}
\hline &$\alpha = 3$ & $\alpha = 3.5 $ & $\alpha = 4 $ &  $\alpha = 4.5 $  & $\alpha = 5 $ &  $\alpha = 5.5 $  \\
\hline$\lambda =0.002 $  & 2.71  & 4.81 &2.74 & 2.41 & 1.63 &1.55  \\
\hline$\lambda =0.004 $  & 1.02 & 0.42  &2.21 & 2.78 & 2.33 &2.29   \\
\hline$\lambda =0.01  $  & 0.37 & 0.41 & 0.04& 3.34  &3.01 &2.97 \\
\hline$\lambda =0.02  $  & 0.15 & 0.18 & 0.08& 0.13 &0.06 &0.06 \\
\hline$\lambda =0.03  $  & 0.09 & 0.12 & 0.09 & 0.21 & 0.11& 0.11 \\
\hline
\end{tabular}
\label{Tab:si_nei_var}
\end{table}

In our simulations, we selected theoretical distance-based and neighbour-based thresholds that could achieve a $95\%$ interference ratio from Table~\ref{Tab:DistanceThreshold} and \ref{Tab:expected95}, and we recorded the variance of the total interference under both types of thresholds. The results for distance-based and neighbour-based thresholds are shown in Table~\ref{Tab:si_dis_var} and Table~\ref{Tab:si_nei_var}, respectively. We observe that distance-based threshold tends to perform worse in low intensity since it has a larger variance. We found out the variance of the neighbour-based threshold is around 100 times smaller than the distance-based threshold in a lower intensity. This is because when we fix the distance-based threshold value for every realisation, there exist some cases where none of the transceivers is within this area. Therefore, distance-based threshold fails to capture the dominant interference. The large variance also affects sum-rate performance which will be verified in Section~\ref{sec:simulation}. 

We also observed that neighbour-based thresholds perform relatively well in both low and high intensity. The neighbour-based threshold is preferable in wireless network optimisation because it guarantees that each target node will be able to connect with at least one neighbouring node. Therefore, neighbour-based threshold is preferable at the intensity of interest.

\subsection{Threshold-based Graph Neural Networks}
In this subsection, we introduce graph representation of both distance-based and neighbour-based methods and the structure of proposed GNNs.
\subsubsection{Graph Representation}
We define the set of vertices and edges of a graph $G$ as $\mathcal{V}$ and $\mathcal{E}$, respectively. For any given vertex $v \in \mathcal{V}$, its set of neighbours is defined as $\mathcal{N}(v)$.

Let $V_v$ and $E_{v,u}$ represent vertex features of vertex $v$ and edge features between vertex $v$ and vertex $u$, respectively. 

With definitions in place, we define the vertex features of the vertices to be
\begin{equation}
V_{v}= \{ h_{ v,D(v)} , w_{v}, d_{ v,D(v)}\} 
\end{equation}
where $h_{v,u} \in \mathbb{C}$ and $d_{v,u} \in \mathbb{R}$ are the channel coefficient and distance between $v$-th transmitter and $u$-th receiver, $w_v$ is the weight for $v$-th transmitter. 
We define the edge features to be
\begin{equation}
E_{v,u}= \{ h_{v,u}, d_{ v,u}\}
\end{equation}
\begin{figure}[H]
\centerline{\includegraphics[width=6cm]{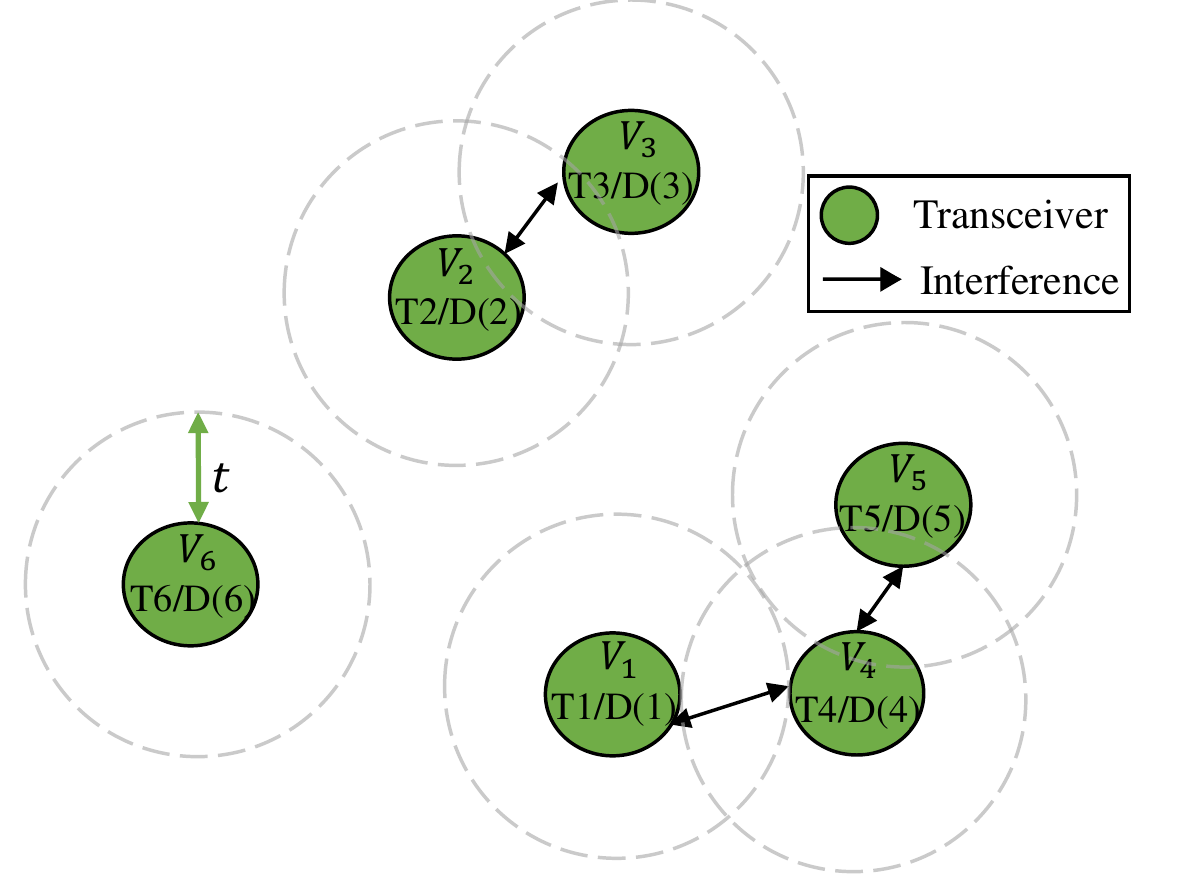}}
\caption{D2D Communication Network with distance-based threshold.}
\label{fig:D2Ddistance}
\end{figure}
In the distance-based threshold method, $v$ only connects $u$ when the distance between them $d_{v,u} $ is smaller than $t(\alpha)$, where $t(\alpha)$ is the appropriate distance-based threshold with path-loss exponent $\alpha$. For example, as in Figure~\ref{fig:D2Ddistance}, the edge between vertex $V_2$ to $V_3$ indicates the distance between receiver $\mathrm{D}(2)$ and transmitter $T_3$ is within the threshold. In the neighbour-based threshold method, $v$ only connects $n$ closest neighbours, where $n(\alpha,\lambda)$ is the appropriate neighbour-based threshold with path-loss exponent $\alpha$ and intensity $\lambda$.

 \subsubsection{The Structure of Graph Neural Network} 
\begin{figure*}[htbp]
\centerline{\includegraphics[width=16cm]{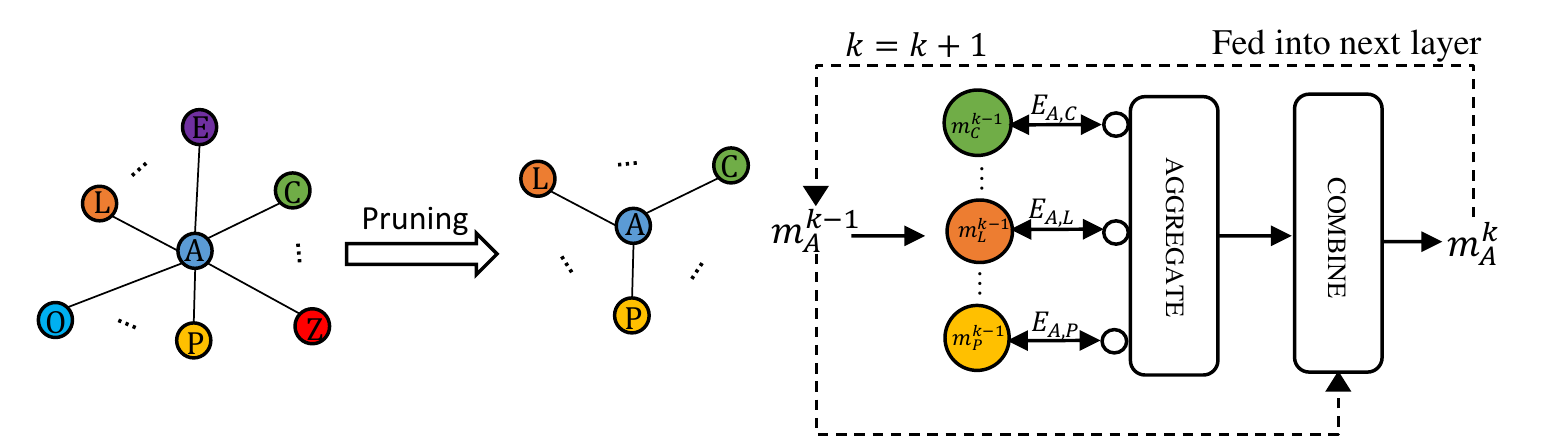}}
\caption{The structure of threshold-based GNN.}
\label{fig:gnnstructure}
\end{figure*}

An illustration of the distance-threshold-based GNN structure is shown in Figure~\ref{fig:gnnstructure}. Our proposed GNNs consist of three steps: Pruning, Aggregation and Combination. First, the edges of a target vertex are pruned based on either distance-based or neighbour-based thresholds. Then, the target vertex collects the information from its current neighbour. We adopt MLP for both aggregating information from a local graph-structured neighbourhood and combining its own features with the aggregated information. Besides, we use the SUM operation to retain the permutation invariance property of GNN.

The updating rule of the proposed threshold-based GNN at $l$-th layer is given by
\begin{equation}
\begin{aligned}
&\text{For distance-based threshold}\\
& \mathcal{N}(v) = \{ u\in \mathcal{V} : d_{v,u} \leq t(\alpha)\}\\
&\text{For neighbour-based threshold}\\
& \mathcal{N}(v) = \{u\in \mathcal{V} : \text{$u$ is the $n(\alpha,\lambda)$ closest neighbours}\} \\
& \alpha_{v}^{(l)}=\operatorname{SUM}\left(\left\{ f_A \left(m_u^{(l-1)}, E_{vu} \right), \forall u \in \mathcal{N}(v)\right\} \right),  \\
& p_{v}^{(l)} =f_C \left(\alpha_{v}^{(l)}, m_{v}^{(l-1)}\right),
\end{aligned}
\end{equation}
where $\alpha_{v}^{(l)}$ and $m_v^{(l)} = \{V_v, p_v^{(l)}\}$ represent the aggregated information from the neighbours and embedding feature vector of vertex $v$, respectively. 

The model includes two 3-layer fully connected neural networks, denoted by $f_{A}$ and $f_{C}$, respectively. Here, $p_v^{(l)}$ represents the allocated power for vertex $v$. 

\section{Simulations and results}\label{sec:simulation}
\subsection{Simulation Setup}
We consider channels with large-scale fading and Rayleigh fading which the CSI is formulated as, $h_{ v,u} = \sqrt{g_{v,u}} r_{v,u}$, where $g_{v,u}= \min \{1, (\frac{d_{v,u}}{d_{0}})^{-\alpha}\}$ $r_{v,u} \sim \mathcal{C}\mathcal{N}\left(0, 1\right)$, $\alpha$ is the path-loss exponent and $d_{0}$ is reference distance.  Here, we consider different intensities within a $100 \times 100$ $m^2$ area and use 1 $m$ as the reference distance\cite{maccartney2014omnidirectional}. Following the similar simulation steps in Section~\ref{subsub:proposedguideline}, we randomly placed the transmitters from $\{20, 40, 100, 200, 300\}$ within a designated area, where the expected intensity $\lambda$ ($T/m^{2}$) ranges from $\{0.002, 0.004, 0.01, 0.02, 0.03\}$. Here, we consider the practical intensities up to 0.03 \cite{AUSTRALIA}. Then each receiver was placed at a random location between $d_{min}=2m$ and $d_{max} = 10m$ away from the corresponding transmitter. Two MLPs $f_A$ and $f_{C}$ with hidden sizes of $\{6,16,32\}$ and $\{36,16,8,1\}$, respectively. We initialise the power $p_{v}^{(0)} = P_{\max}$. To develop an effective power allocation strategy for our wireless network, we adopted the negative sum rate as our objective loss function, which can be expressed in \eqref{eq:lossfunction}. 
\begin{equation}
\begin{aligned}
L(\theta)= & -\hat{\mathbb{E}}_{\boldsymbol{H}}\Bigg\{\sum_{t \in \mathcal{T}} w_{t} \log _{2}(1+\frac{\left|h_{t,\mathrm{D}(t)}\right|^{2} p_{t}}{ \sum_{j\in  \mathcal{T} \backslash\{t\}} \left|h_{j, \mathrm{D}(t)}\right|^{2} p_{j}  +\sigma_{\mathrm{D}(t)}^{2}})\Bigg\} \\
\end{aligned}
\label{eq:lossfunction}
\end{equation}

In practice, we generate 10000 training samples for calculating the empirical loss, and we also generate 2000 testing samples for evaluation. We assumed that only partial CSI (a subset of full CSI) is available to the algorithms.

To validate the effectiveness of our proposed guideline, we conduct the experiments  among the following four algorithms:
\begin{itemize}
    \item WCGCN \cite{shen2020graph}: This is state-of-art GNN algorithm for power allocation problem. It should be noted that the complete graph is used in this approach. 
    \item D-GNN: Our proposed GNN when the distance-based threshold is applied.
    \item N-GNN: Our proposed GNN when the neighbour-based threshold is applied.
    \item WMMSE \cite{shi2011iteratively}: This is the advanced optimisation-based algorithm for power allocation in wireless networks, also see  \cite{sun2018learning, liang2019towards, shen2020graph} for references.
    \item Heuristic: We allocate the maximum power $P_{\text {max}}$ for $0.5T$ pairs which have the $0.5T$th-largest communication channel gain among all $T$ pairs, while the rest are set as 0. 

\end{itemize}
\subsection{Performance}

 \begin{figure}[htbp]
\centerline{\includegraphics[width=8cm]{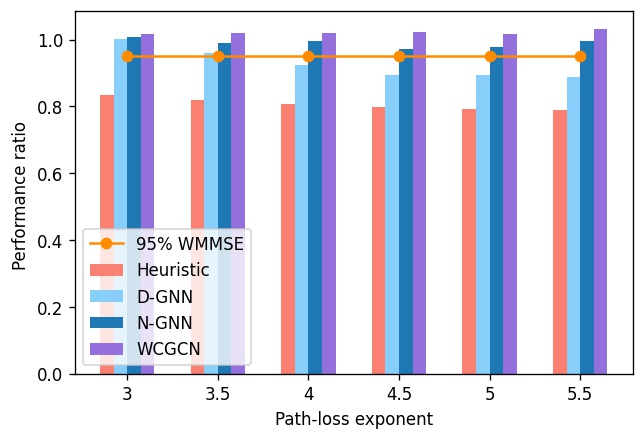}}
\caption{Normalized average weighted sum rate with thresholds that achieve $95\%$ interference (Tables \ref{Tab:DistanceThreshold}  and \ref{Tab:expected95}) under $\lambda=0.01$.}
\label{fig:performance_all}
\end{figure}
We conducted an investigation into the effectiveness of the proposed guideline by setting the threshold to the values that capture $95\%$ interference from Table~\ref{Tab:DistanceThreshold} and \ref{Tab:expected95} for both D-GNN and N-GNN. The sum-rate performance of all the algorithms under the same density $\lambda = 0.01$ is presented in Figure~\ref{fig:performance_all}. Since WMMSE is the state-of-art optimisation-based algorithm in power allocation problems, we normalise all the performance by the performance of WMMSE under the same setup for the rest of the experiments unless specified. The results indicate that N-GNN tends to perform better than D-GNN in all scenarios due to its flexibility in using neighbour-based threshold. We further test the performance of the proposed algorithm under the density $\lambda = 0.004$ and path-loss exponent $\alpha=3.5$, but different $d_{min}$ and $d_{max}$ is presented in Table~\ref{Tab:GNN_[1,5]}. When assessing performance across various distance distributions, N-GNN showcases a solid ability to deliver satisfactory results. However, D-GNN approach lags behind other GNN-based methods in terms of performance. This discrepancy can be attributed to the larger variance, which is discussed in Section~\ref{subsub:proposedguideline}. It is noteworthy that WCGCN achieves better performance by using all the available information, but it comes at the cost of increased time complexity as will be discussed in Section~\ref{sec:timecomplexity}.  
\begin{table}
\centering
\caption{Normalised performance with different distance distribution under $\lambda = 0.004$ and $\alpha=3.5$}
\begin{tabular}{|c|c|c|c|c|}
\hline$[d_{min},d_{max}]m$  & $[2,20]$ & $[5,5] $  & $[5,15]$& $[10,30]$      \\
\hline Heuristic & $75.0\%$ &  $75.2\%$ & $70.1\%$ & $53.6\%$     \\
\hline WCGCN &  $101.7\%$& $100.1\%$  &$100.9\%$  & $91.0\%$    \\
\hline N-GNN  & $96.7\%$ &  $97.8\%$  & $96.4\%$ & $87.5\%$  \\
\hline D-GNN&  $84.2\%$ & $88.0\%$  & $82.9\%$ & $82.1\%$   \\
\hline
\end{tabular}
\label{Tab:GNN_[1,5]}
\end{table}

\subsection{Generalisation}
\subsubsection{Generalisation to Varied Spatial Dimensions}
We first train N-GNN and D-GNN with 40 transceiver pairs within a $100\times 100m$ region, then we test the trained networks with different sizes of the region but under the same intensity ($\lambda = 0.004$). The results are demonstrated in Table~\ref{Tab:GNN_generalistion_size}.  Evidently, it is observed that the performance remains consistent despite alterations in spatial dimensions. This observation highlights the capacity of our proposed algorithms to effectively generalise across different scenarios.
\begin{table}[H]
\centering
\caption{Generalise to Varied Spatial Dimensions ($\alpha =3.5$, $\lambda=0.004$).}
\begin{tabular}{|c|c|c|c|}
\hline Transceiver Pairs  & Size($m^2$) & N-GNN  & D-GNN    \\
\hline $10$ & $50\times50$ &  $100.3\%$ & $94.3\%$    \\
\hline $160$ & $200\times200$ & $100.2\%$  & $95.8\%$    \\
\hline $360$ & $300\times300$ & $100.1\%$  & $95.9\%$    \\
\hline $640$ & $400\times400$ & $100.1\%$  & $95.8\%$    \\
\hline
\end{tabular}
\label{Tab:GNN_generalistion_size}
\end{table}
\subsubsection{Generalisation to Varied Network Densities}
Similarly, we first train N-GNN and D-GNN with 100 transceiver pairs in a $100\times 100m$ area, then we test the trained networks with varied transceiver pair quantities while keeping the region dimension consistent. The outcomes are provided in Table~\ref{Tab:GNN_generalistion_density}. Notably, our proposed N-GNN achieve an impressive $98\%$ of the state-of-the-art WMMSE algorithm's performance, even when faced with a quadrupled density increase. This indicates the adaptability of our proposed algorithms to navigate complex network environments.
\begin{table}
\centering
\caption{Generalise to Varied Network Densities($\alpha =3.5$).}
\begin{tabular}{|c|c|c|c|}
\hline Transceiver Pairs  & Size($m^2$) & N-GNN  & D-GNN    \\
\hline $20$ & $100\times100$ &  $99.3\%$ &  $94.3\%$   \\
\hline $40$ & $100\times100$ &  $99.9\%$ &  $95.1\%$     \\
\hline $200$ & $100\times100$ &  $98.5\%$ &  $96.0\%$      \\
\hline $300$ & $100\times100$ & $98.4\%$ &  $96.3\%$     \\
\hline $400$ & $100\times100$ & $97.9\%$ &  $96.6\%$     \\
\hline
\end{tabular}
\label{Tab:GNN_generalistion_density}
\end{table}
\subsection{Time Complexity}\label{sec:timecomplexity}
\begin{figure}[htbp]
\centerline{\includegraphics[width=8cm]{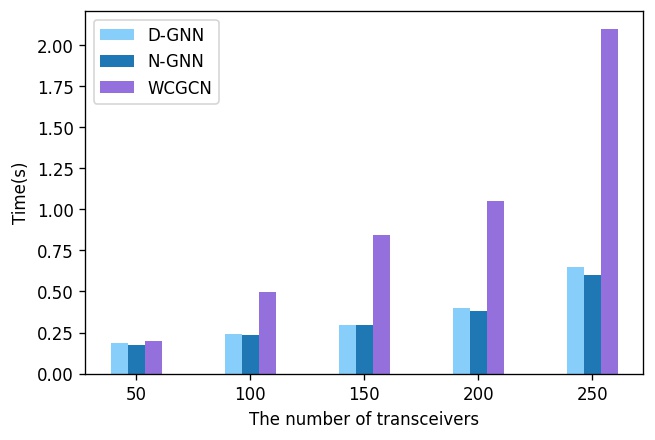}}
\caption{Average inference time with thresholds that achieve $95\%$ interference (Tables \ref{Tab:DistanceThreshold}  and \ref{Tab:expected95}) under $\alpha =5.5$ and $\lambda=0.01$.}
\label{fig:timecomplexity_all}
\end{figure}

\begin{figure}[htbp]
\centerline{\includegraphics[width=8cm]{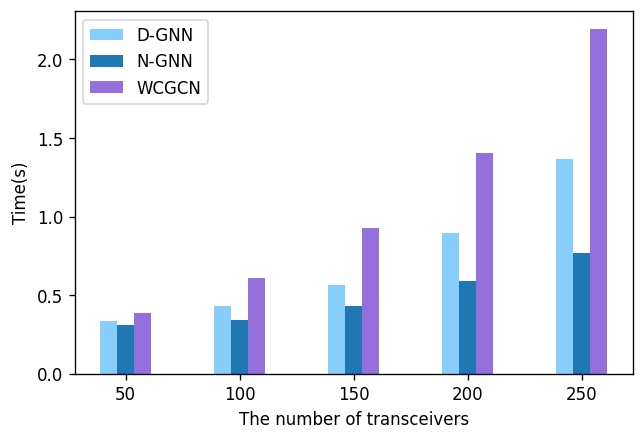}}
\caption{Average inference time for all algorithms to achieve $98\%$ performance of WMMSE ($\alpha =5.5$, $\lambda=0.01$).}
\label{fig:timecomplexity_98}
\end{figure}

The average inference time for the algorithms under the same experimental setting as Figure~\ref{fig:performance_all} is depicted in Figure~\ref{fig:timecomplexity_all}. We observe that N-GNN  has significantly lower complexity compared to WCGCN. For instance, by selecting $|V| =200$, N-GNN achieves a $95\%$ optimal performance while reducing the required inference time by roughly $63\%$ compared to WCGCN. To further investigate the time complexity of N-GNN and D-GNN, the inference time for all algorithms when the performance achieves $98\%$ of WMMSE is shown in Figure~\ref{fig:timecomplexity_98}. We observe that D-GNN requires around $40\%$ time more than N-GNN when $T=250$. Furthermore, the observed time disparities are amplified with an increase in the number of transmitters.

This is due to the fact that the time complexity for each GNN layer is $\mathcal{O}(|\mathcal{V}|+|\mathcal{E}|)$, which is dominated by $|\mathcal{E}|$, and pruning the edges helps to reduce the complexity. In WCGCN and D-GNN, the time complexity is $\mathcal{O}(|\mathcal{V}|^2)$ \cite{shen2020graph,chen2023graph}, while in N-GNN, the total number of edges for each transceiver pair is fixed by the threshold $n(\alpha,\lambda)$. Since the total number of edges $|\mathcal{E}| = n(\alpha,\lambda) |\mathcal{V}|$, the time complexity for neighbour-based GNN will be $\mathcal{O}((1+n(\alpha,\lambda))|\mathcal{V}|)$. Therefore, we could reduce the time complexity from quadratic to linear with the number of transceiver pairs by introducing the neighbour-based threshold.  

\section{Conclusion}
This research provides recommendations for selecting the appropriate threshold value in terms of their potential to catch the expected interference. We show that by choosing a suitable neighbour-based threshold, our proposed neighbour-based GNN can preserve strong performance and generalisation capacity, while the time complexity is reduced from $\mathcal{O}(|\mathcal{V}|^2)$ to $\mathcal{O}(|\mathcal{V}|)$, where $|\mathcal{V}|$ is the total number of transceiver pairs.

\bibliographystyle{IEEEtran}
\bibliography{main}

\vspace{12pt}

\end{document}